\documentclass[aps,prb,reprint,superscriptaddress,twocolumn,showpacs]{revtex4-2}
\usepackage{float}
\usepackage[french,english]{babel}
\usepackage[T1]{fontenc}
\usepackage[utf8]{inputenc}

\usepackage{graphics,graphicx,caption,subcaption}
\usepackage[dvipsnames,usenames]{color}
\usepackage{rotating}
\usepackage{epstopdf}

\usepackage{amssymb,amsmath}
\usepackage{xfrac,nicefrac}
\usepackage{gensymb}
\usepackage{bm}
\usepackage{pst-node}
\usepackage{relsize}
\usepackage{units}

\usepackage[active]{srcltx}
\usepackage[sort&compress]{natbib}

\usepackage{hyperref}

\usepackage{array}
\usepackage{tabularx}
\usepackage{booktabs}
\usepackage{multirow} 
\usepackage{dcolumn}

\usepackage[color=orange!20]{todonotes} 
\usepackage[normalem]{ulem}

\newcolumntype{d}[1]{D{.}{.}{#1}}

\newlength{\dbarheight}

\newlength{\ddbarheight}

\begin{document}

\date{February 2026}

\title{\textit{Ab-initio} study of magnetic interaction in RNiO$_2$: role of
  rare-earth size and epitaxial strain}

\author{Ramasamy MURUGESAN}
\affiliation{
 Institut Néel CNRS, 38042 Grenoble, France.
}

\author{Shiyu DENG}
\affiliation{
Institut Laue Langevin, 38042 Grenoble, France.
}

\author{Marie-Bernadette LEPETIT}
 \email{Contact author: Marie-Bernadette.Lepetit@neel.cnrs.fr}
\affiliation{
 Institut Néel CNRS, 38042 Grenoble, France.
}
\affiliation{
Institut Laue Langevin, 38042 Grenoble, France.
}

\begin{abstract}

  In this paper we present a systematic first-principles investigation of the
  structure and nearest-neighbor magnetic exchange interactions in the
  infinite-layer RNiO$_2$ thin-film series (R $=$ La–Lu), under the constrain
  of different substrates. Density functional theory calculations are first
  employed to optimize the crystal structures, while CAS+DDCI correlated
  wave-function calculations were used for the determination of the magnetic
  exchanges. This methodology allows to obtain magnetic interactions in good
  agreement with RIXS experimental estimations of
  Ref.~\onlinecite{lu2021magnetic, gao2024magnetic,PhysRevB.109.024512}. In
  the pristine case, our results confirm that the  structure of
  RNiO$_2$ (except for R $=$ La and Ce) adopts the ${I4/mcm}$ space group,
  characterized by an out-of-plane rotation of the NiO$_4$ plaquettes. This
  result remains valid under substrate constrains, even if compressive strain
  reduces the amplitude of the rotational distortion and leads to an
  enhancement of the magnetic exchange interaction relative to the pristine
  structures. The latter are found to depend sensitively on both the
  rare-earth ionic radius and the applied epitaxial strain.  These results
  highlight the role of structural distortions and substrate-induced strain in
  tuning magnetic interactions in infinite-layer nickelates.

\end{abstract}

\maketitle

\section{Introduction}

The discovery of superconductivity in infinite–layer NdNiO$_2$~\cite{ILNickel} has revived
strong interest in nickelate–based systems as potential analogs of the cuprate
high–temperature superconductors~\cite{Cu_HighTC}. Because Ni lies
next to Cu in the periodic table, nickelates were among the earliest candidate
materials considered after the discovery of superconductivity in
cuprates. However, superconductivity in nickelates was only realized recently
(2019) in thin films of infinite–layer NdNiO$_2$, with critical
temperature between \unit[9-15]{K}~\cite{ILNickel}. Beyond the search for materials
with higher critical temperatures, studies of such analog systems aim to
clarify the fundamental mechanisms responsible for electron pairing in
unconventional superconductors. 

Nickelates occupy a particularly intriguing position in this context. In the
infinite–layer phase, both nickelates and cuprates exhibit a nominal $3d^9$
electronic configuration, corresponding to Ni$^{1+}$ in nickelates and
Cu$^{2+}$ in cuprates. In addition, both systems feature square–planar
coordination of the transition–metal ion by oxygen, and display
superconducting domes upon hole doping. 

Despite these similarities, important differences exist between the two
material families. The parent cuprates present well–established long–range antiferromagnetic order~\cite{ZhangRice,Cu_Mag,Cu_Mag2} whereas infinite–layer nickelates do not
exhibit long–range magnetic order~\cite{NiPES}. Instead, experiments suggest
the presence of predominantly short–range magnetic
correlations~\cite{lu2021magnetic,gao2024magnetic}, as expected in a 2D
system.

One of the most striking difference between nickelates and cuprate is however
the recent discovery of superconductivity in the parent, unstopped nickelates
films. Indeed, very pure thin films of NdNiO$_2$~\cite{ILNickel,
  parzyck2025superconductivity} and
PrNiO$_2$ on SrTiO$_3$ substrates~\cite{sahib2025superconductivity} and
  more recently of SmNiO$_2$ on (LaAlO$_3$)$_{0.3}$(Sr$_2$TaAlO$_6$)$_{0.7}$
  (LSAT) substrates~\cite{yang2026enhanced} were recently reported to be
superconducting below \unit[11]{K}.

(Eu,Ca) Co-doped samarium nickelate achieve a record ~\cite{yang2026enhanced}, as they reach the smallest $c$-axis parameter (\unit[3.26]{\AA})
and the largest superconducting transition temperature (\unit[32.5]{K}), revealing a clear correlation
between decreasing $c$-axis parameter and increasing critical
temperature.

Another distinctive aspect of the infinite–layer nickelates is the role of the 
rare–earth ions. In cuprates, rare–earth elements primarily act as spacer
layers that separate the CuO$_2$ planes. In contrast, DFT calculations see
nickelates exhibiting a more complex multi-band electronic structure, that
includes contributions from rare–earth states near the Fermi level~\cite{PhysRevB.100.201106, PhysRevX.11.011050, PhysRevB.100.205138}, which are
largely absent in the cuprates. 

In unconventional superconductors such as the cuprates, antiferromagnetic (AFM) spin
fluctuations are widely believed to play a key role in the pairing
mechanism. Understanding the magnetic exchange interactions in the parent
compounds of the nickelates is therefore essential for elucidating the origin
of superconductivity in these systems. Although long–range magnetic ordering
has not been experimentally observed in RNiO$_2$, resonant inelastic X–ray
scattering (RIXS) measurements on NdNiO$_2$~\cite{lu2021magnetic} and
PrNiO$_2$~\cite{gao2024magnetic,PhysRevB.109.024512} thin films, grown on SrTiO$_3$ substrates, have
revealed dispersive spin excitations with an estimated AFM nearest–neighbor
exchange interaction of approximately \unit[-60]{meV}.

A further remarkable feature of infinite–layer nickelates is that
superconductivity has only been observed in thin–film form. Attempts
to realize superconductivity in bulk infinite–layer nickelates have so far
been unsuccessful~\cite{li2020absence}. This raises important questions
regarding the role of substrate–induced epitaxial strain in stabilizing
superconductivity in these materials. Strain engineering is known to influence
superconducting properties in other unconventional superconductors, including
the cuprates, where compressive strain has been associated with enhanced
superconducting transition temperatures through modifications of the magnetic
exchange interactions~\cite{locquet1998doubling, sato1997increase,
  ivashko2019strain}. Experimental studies on Sr–doped PrNiO$_2$ films have
likewise reported variations in the superconducting critical temperature
depending on the choice of substrate~\cite{ren2023possible}.

Another important aspect is the effect of rare-earth nature. Indeed, recent theoretical
studies have suggested that structural distortions involving rotations of the
NiO$_4$ units may occur as the rare–earth ionic size decreases
\cite{philipghosez,xia,carrasco}, which would modify the Ni--O--Ni bond angle and
consequently the magnetic superexchange interaction. 

These observations motivate a systematic investigation of how structural
parameters influence magnetic interactions in RNiO$_2$. In particular,
understanding how the magnetic exchange interaction varies with rare–earth
size and epitaxial strain may provide important insight into the magnetic
correlations present in these systems.

In this work, we perform a comprehensive theoretical study of the entire
rare–earth series in infinite–layer RNiO$_2$ (R $=$ La–Lu) with the aim of
determining the nearest–neighbor magnetic exchange interaction $J$. First,
density functional theory (DFT) calculations are employed to optimize the
crystal structures and to examine the dependence of lattice parameters and
structural stability on the rare–earth ionic radius. To account for
experimentally relevant conditions, we also consider the effect of epitaxial
strain by constraining the in–plane lattice parameters to those corresponding
to common substrates such as LaAlO$_3$ and SrTiO$_3$. Finally, the magnetic
exchange interactions for both pristine and strained structures are computed
so that to provide a detailed theoretical understanding of how rare–earth and
substrate strain influence magnetic interactions in the RNiO$_2$ family. As
the infinite-layer nickelates are strongly correlated~\cite{katukuri}, we used
for this purpose a multireference wave-function–based quantum chemical method,
namely the Complete Active Space + Difference Dedicated Configuration
Interaction~\cite{DDCI1992,IDDCI1995} method, that was specifically designed
to accurately compute excitation energies in strongly correlated compounds.

\section{Methods}

The geometric structure of RNiO$_2$ was obtained through DFT-based structural
optimization using the CRYSTAL23 code~\cite{crystal23}. To identify the
lowest-energy configuration, full geometry optimizations were performed in the
${P4/mmm}$ and ${I4/mcm}$ space groups. Magnetic ordering was taken into
account by employing a $\sqrt{2}a \times \sqrt{2}a \times 2c$ supercell with C
and G-type (for a few test cases) antiferromagnetic ordering. No significative
difference in geometries were found on the tested cases between the C- and
G-orderings. Ni and O atoms were described with an all-electron valence
$3\zeta + p$ basis set~\cite{peintinger2013consistent, pernot2015prediction},
whereas the rare-earth elements were treated using Stuttgart
scalar-relativistic core pseudopotentials~\cite{dolg1989pseudopotential,
  dolg1989energy}. Exchange–correlation effects were approximated using the
WC1LYP functional~\cite{demichelis2010performance}. The functional choice was
motivated by it's ability to reproduce the experimental structure of bulk
NdNiO$_3$~\cite{RNiO3_Struc1,RNiO3_Struc2} and infinite layer
NdNiO$_2$~\cite{HAYWARD2003839} with 1.1~\% error margin.  Further, a
$16 \times 16 \times 16$ Monkhorst-Pack $k$-point grid and a Gilat grid of
$32 \times 32 \times 32$ were used for density matrix evaluation. Structural
relaxations were performed until the interatomic forces were below
\unit[12.25]{meV/\AA}. Substrate-induced strain was simulated by performing
constrained structural relaxations in which the in-plane lattice parameters
were fixed to those of LaAlO$_3$ (\unit[$a=b=5.366$]{\AA}) and SrTiO$_3$
(\unit[$a=b=5.5134$]{\AA}), while allowing the atomic positions and the
out-of-plane lattice parameter to relax.

To accurately determine the magnetic exchange interactions for the
DFT-optimized structures, multireference configuration interaction methods
were employed. In particular, the complete active space plus single and double
excitations of the difference dedicated configuration interaction (CAS+DDCI)
approach~\cite{DDCI1992,IDDCI1995} was used.  The CAS+DDCI method was
specifically designed to accurately obtain energy differences between magnetic
states based on the same orbital space. It includes both the exact treatment
of the electronic correlation between the magnetic orbitals, and the
differential screening effects on each of the previous configurations up to
two-holes
two particles and their interactions.\\
As in reference~\onlinecite{katukuri} the calculations were performed on an
embedded fragment based on the two Ni atoms involved in the sought $J$ and
their first coordination shell (see Supplementary Information for
pictures~\cite{SI}).  This fragment is embedded in a larger cluster of
self-consistently optimized Total Ions
Pseudopotentials~\cite{TIP,EmbedAIMP_Seijo,Scepic_2022}
(TIPs).  Finally a larger set of renormalized point charges is defined in
order to obtain an exponential convergence of the Madelung potential series
following the method of reference~\onlinecite{p:env2}.  Fragment orbitals were
optimized, using the Ni$^{1+}$-ions magnetic-orbitals as active space, at the
complete active space self-consistent field (CASSCF) level using the MOLCAS
package~\cite{MOLCAS_8}.

Recent ARPES measurements~\cite{ding2024} on LaNiO$_2$ indicate the existence
of an electron pocket around the $A$ point of the Fermi surface, more recently
C. Li {\it et al} ARPES measurements~\cite{arXiv:2507.04378} clarified the
nature of this electron pocket as predominantly of interstitial-$s$ character,
located between adjacent Ni atoms along the $c$-direction, with negligible
rare-earth $5d$ and $4f$ contributions.  We thus ensured that such
interstitial orbitals can be accounted for within our basis set. For this
purpose we use an all-electrons, $3\zeta+p$ basis sets~\cite{roos2005new,
  roos2004main} with diffuse enough orbitals (see supplementary information
for a picture~\cite{SI}).

The resulting optimized orbitals and integrals were subsequently used in
CAS+DDCI calculations, carried out with the RelaxSE code~\cite{RelaxSE}, to
obtain the low-energy states of the fragment. The magnetic exchange
interactions were obtained by mapping the computed low-energy excitation energies onto
a Heisenberg Hamiltonian,
\begin{equation}
    \hat{H} = - \sum_{i, j} J_{ij} \hat{\mathbf{S}}_i \cdot \hat{\mathbf{S}}_j,
\end{equation}
where the sum runs over nearest neighbor Ni-Ni bonds, $\hat{\mathbf{S}}_i$ and
$\hat{\mathbf{S}}_j$ are the $\sfrac{1}{2}$ quantum spin operators associated
with sites $i$ and $j$. As for the cuprates, the underlying magnetic orbitals
are mainly supported by the $d_{x^2-y^2}$ orbitals of the Ni sites with some
delocalization over the neighboring 2p orbitals of the surrounding oxygen (see
supplementary information~\cite{SI} for a picture of the {\it ab-initio}
computed magnetic orbitals). In this framework the $J_{ij}$ exchange
interactions are positive for ferromagnetic (FM) interactions and negative for
AFM ones.

\section{Results and Discussion}

\subsection{Structural Optimization} 
\begin{figure*}[htbp]
  \centering
  \begin{subfigure}{0.49\textwidth}
    \centering
    \includegraphics[width=\linewidth, height=2.5in, keepaspectratio]{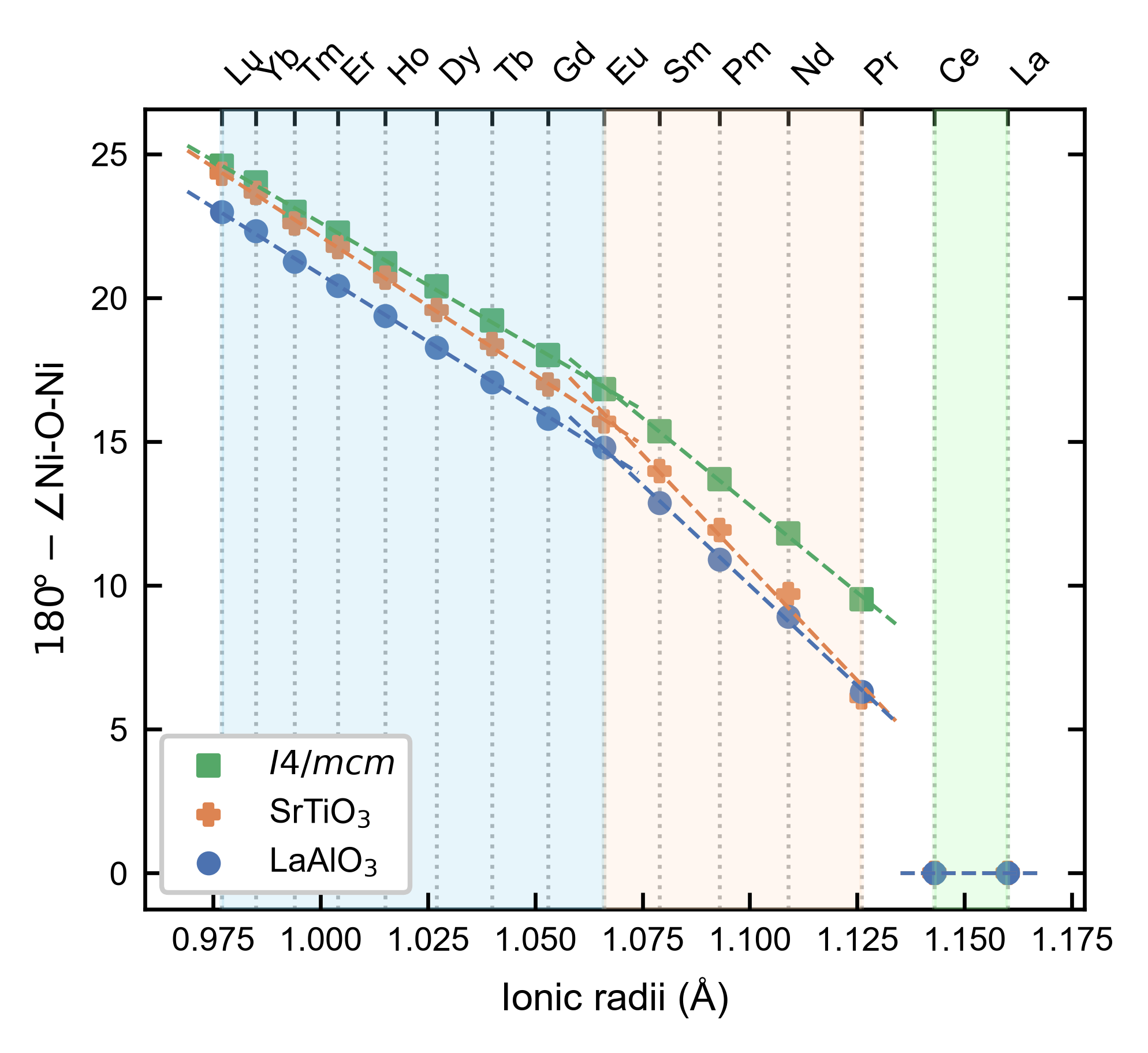}
    \caption{}
    \label{fig:a}
  \end{subfigure}
  \hspace{-0.9em}
  \begin{subfigure}{0.49\textwidth}
    \centering
    \includegraphics[width=\linewidth, height=2.5in, keepaspectratio]{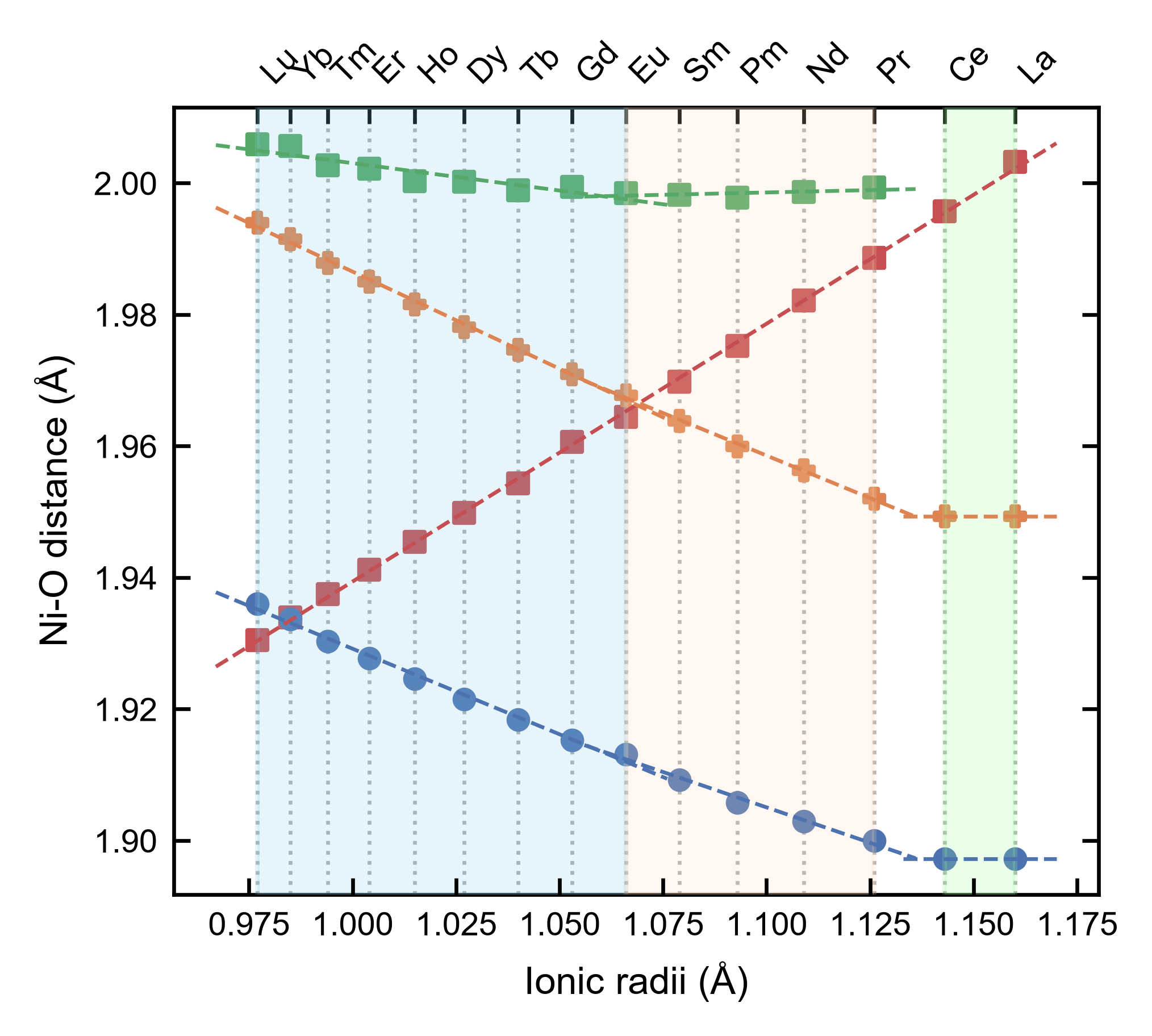}
    \caption{}
    \label{fig:b}
  \end{subfigure}
  \begin{subfigure}{0.49\textwidth}
    \centering
    \includegraphics[width=\linewidth, height=2.5in, keepaspectratio]{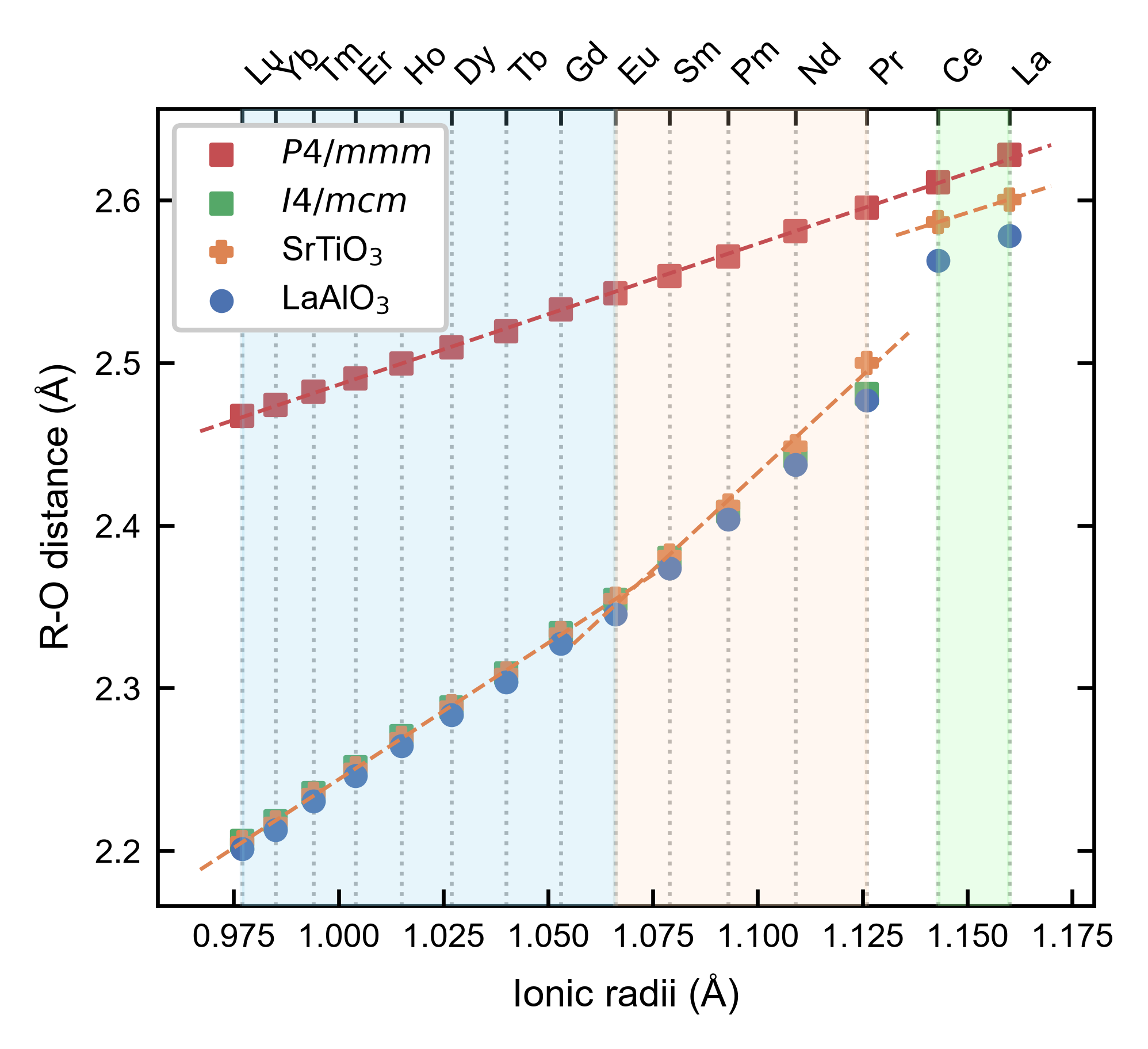}
    \caption{}
    \label{fig:c}
  \end{subfigure}
  \hspace{-0.9em}
  \begin{subfigure}{0.49\textwidth}
    \centering
    \includegraphics[width=\linewidth, height=3.5in, keepaspectratio]{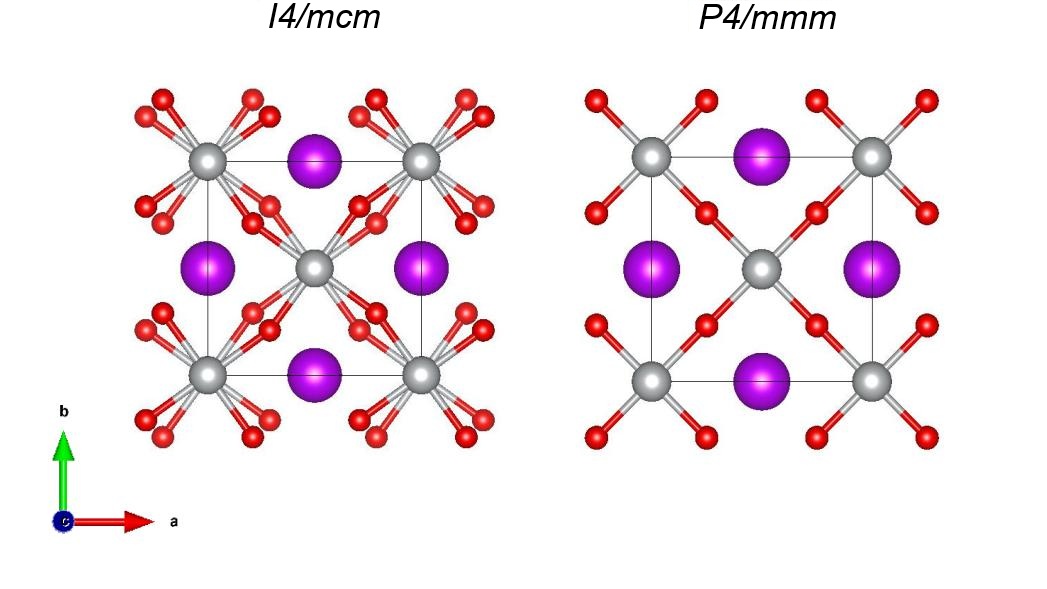}
    \caption{}
    \label{fig:d}
  \end{subfigure}
  \caption{(a) Ni--O--Ni bond angle, (b) Ni--O bond length, and (d)
    shortest R--O bond length of RNiO$_2$ (R $=$ La-Lu) in the bulk and
    under substrate strain. Linear fits (dashed lines) are applied to three
    subspaces: Lu--Gd (blue shaded region), Eu--Pr (orange shaded region), and
    Ce--La (green shaded region), indicating the ionic-radius range of each
    subspace. (c) Optimized representative geometries of RNiO$_2$ in the
    ${I4/mcm}$ and ${P4/mmm}$ phases, where purple, gray, and red denote R,
    Ni, and O atoms, respectively.}
  \label{fig:StrucCompare}
\end{figure*}

The RNiO$_2$ compounds, particularly NdNiO$_2$, have predominantly been
investigated within the high-symmetry ${P4/mmm}$ space group. In
contrast, their perovskite counterparts RNiO$_3$ are well known to undergo
structural distortions involving rotations of the NiO$_6$ octahedra, with the
rotation amplitude increasing systematically as the rare-earth ionic radius
decreases~\cite{NiO6Angle}. A similar tendency toward rotational distortions has been suggested
for RNiO$_2$. Previous first-principles calculations~\cite{philipghosez,xia,carrasco}
identified unstable phonon modes associated with out-of-plane rotations of the
NiO$_4$ square-planar units in AFM configurations. For smaller
rare-earth ions, a dynamical instability corresponding to the $A^-_4$ mode was
reported~\cite{philipghosez}, indicating that the energetically preferred
structure may adopt the lower-symmetry ${I4/mcm}$ phase characterized by
an $a^{0}a^{0}c^{-}$ rotational pattern.

To examine this possibility, full structural optimizations were performed
using a hybrid exchange-correlation functional, initializing the calculations
from both the high-symmetry ${P4/mmm}$ phase (without rotations) and the
distorted ${I4/mcm}$ phase (with finite $a^{0}a^{0}c^{-}$ rotations). The
computed total energy differences (provided in supplementary
  information~\cite{SI}) indicate that the ${I4/mcm}$ structure is energetically
favored for all RNiO$_2$ compounds except LaNiO$_2$ and CeNiO$_2$. In the
${P4/mmm}$ phase, the Ni--O--Ni bond angle is linear, however in the
${I4/mcm}$ phase, the Ni--O--Ni bond angle is distorted, deviating from
$180^{\circ}$ as shown in Fig.~\ref{fig:a}. This distortion is further
influenced by the choice of substrates, with variations observed across
different substrates.

The transition from a linear to a distorted Ni--O--Ni geometry indicates a
structural relaxation driven by ionic size effects, which stabilizes the
${I4/mcm}$ phase over ${P4/mmm}$. As shown in Fig.~\ref{fig:a}, the Ni--O--Ni
angle displays three distinct trends (dashed lines) corresponding to different
rare-earth groups: Lu-Gd and Eu-Pr exhibit different slopes for the rotation
angle as a function of the rare-earth ionic-radius (indicated by different
shading), while La- and Ce-based compounds adopt a linear Ni--O--Ni
configuration (highlighted with a distinct shade). A schematic representation
of the optimized RNiO$_2$ geometry in the ${I4/mcm}$ and ${P4/mmm}$ phases is
provided in Fig.~\ref{fig:d}, where one can clearly visualize the linear and
distorted Ni--O--Ni bond angles. It should be noted that Ni--O--Ni angles
exhibit a non-continuous behavior at the transition located between the Pr and
Ce compounds. This points toward a first-order transition despite the
group/subgroup relationship between $P4/mmm$ and $I4/mcm$.  The boundary
between these phases corresponds to \unit[$R_{\text{ion}}\simeq
1.14$]{\AA}. It slightly differs from that reported in other first-principles
studies~\cite{xia,carrasco}. In particular, Xia \textit{et al.}~\cite{xia}
obtained a more extended stability range of the ${P4/mmm}$ phase with
SmNiO$_2$ identified transition boundary. This difference can be attributed to
the characteristics of the chosen exchange-correlation functionals. For
instance the PBE and PBESOL functionals, used by Xia \textit{et al.}, are
known to over-delocalize and thus favor the most symmetric structures compared
to hybrid functionals.

With decreasing rare–earth ionic radius from La to Lu, the unit–cell volume
progressively contracts, resulting in a reduction of the in–plane Ni--Ni
distance. In the high-symmetry ${P4/mmm}$ structure, the oxygen atom resides
at the midpoint of the Ni--Ni bond, enforcing a linear Ni--O--Ni bond angle of
$180^{\circ}$. For LaNiO$_2$ and CeNiO$_2$, the relaxed Ni--O bond length in
${P4/mmm}$ is approximately \unit[1.99]{\AA}. If the smaller rare–earth
compounds were constrained to remain in this high-symmetry phase, the reduced
lattice parameter would further shorten the Ni--O bond (down to \unit[$\sim$
1.93–1.98]{\AA}), leading to significant bond compression. Instead, the system
accommodates the lattice contraction through an $a^{0}a^{0}c^{-}$ rotational
distortion of the NiO$_4$ squares, lowering the symmetry to ${I4/mcm}$.

This distortion preserves a nearly constant Ni--O bond length (\unit[$\sim$
1.99-2.00]{\AA}) (Fig.~\ref{fig:b} green markers) while reducing the
Ni--O--Ni bond angle below $180^{\circ}$ (Fig.~\ref{fig:a}). The
deviation from linearity increases systematically from Pr to Lu. The
stabilization of the ${I4/mcm}$ phase can therefore be attributed to the
energetic gain associated with avoiding excessive Ni--O bond compression. A
summary of the optimized structural parameters for both space groups is
provided in the Supplementary Information~\cite{SI}.

Let us now examine the effect of epitaxial strain, by performing constrained
structural relaxations where the in–plane lattice constants are fixed to those of the LaAlO$_3$ and SrTiO$_3$
substrates. Relative to the fully relaxed
${I4/mcm}$ structures, these substrates impose compressive strain. Under
these constraints, the out-of-plane rotational distortion remains present for
all rare–earth ions (except LaNiO$_2$ and CeNiO$_2$, which remain in the
${P4/mmm}$ phase, even under compressive strain), and its amplitude generally increases with decreasing ionic radius as
shown in Fig.~\ref{fig:a}.

For a given rare–earth ion, however, the rotation angle is slightly reduced
under compressive strain compared to the fully relaxed bulk ${I4/mcm}$
structure. Because the in–plane lattice constant is fixed by the substrate,
the Ni--Ni distance becomes strain–controlled, limiting the extent to which
the NiO$_4$ squares can rotate to maintain the optimal Ni--O bond length. In
this strained regime, the geometry is additionally influenced by the
shortest R--O distance, which remains nearly identical to its value in
the unconstrained ${I4/mcm}$ phase, and thus independent of the imposed
constrain (Fig.~\ref{fig:c}), except for the LaNiO$_2$ and CeNiO$_2$ systems.
This behavior indicates that the shortest R--O distance acts as an important geometric
constraint, indirectly influencing the Ni--O--Ni bond angle. Since this bond
angle strongly affects the super-exchange pathways, it is expected to play a
central role in the determination of the magnetic interactions, as discussed in the
following section.

\subsection{Magnetic Exchange Interaction} 

The nearest–neighbor magnetic exchange interactions between Ni ions were 
evaluated using \textit{ab-initio} calculations along the in-plane and
out-of-plane direction. In RNiO$_2$, Ni formally adopts a $+1$ oxidation state
with a $3d^9$ electronic configuration. Under square–planar coordination, the
magnetic orbital is predominantly of $d_{x^2-y^2}$ character. The magnetic
coupling between adjacent Ni sites along the in-plane direction is therefore
mediated by the oxygen $p_{\sigma}$ orbitals via a super-exchange
mechanism. The calculated exchange constants $J_{\text{in}}$ are summarized in
Table~\ref{tab:table2}. For all compounds and structural configurations
considered in this work, the nearest–neighbor interaction is found to be
antiferromagnetic.

\begin{table}[htbp]
  \caption{ In-plane magnetic exchange interaction $J_{\text{in}}$ (in meV)
    obtained from \textit{ab-initio} calculations for the RNiO$_2$
    family. Different space groups and epitaxial strains (SrTiO$_3$ and
    LaAlO$_3$ substrates) are reported.}
\label{tab:table2}
\begin{ruledtabular}
\begin{tabular}{ccccc}
  System & \multicolumn{2}{c}{Bulk phase} & \multicolumn{2}{c}{Thin films} \\
  & ${P4/mmm}$ & ${I/4mcm}$ & SrTiO$_3$ & LaAlO$_3$ \\
 &  &  &  &  \\
\hline
LaNiO$_2$ & -48.3 & - & -72.4 & -106.0 \\
CeNiO$_2$ & -49.5 & - & -69.8 & -101.9 \\
PrNiO$_2$ & -50.7 & -46.1 & -65.9 & -96.5 \\
NdNiO$_2$ & -51.9 & -45.0 & -61.8 & -91.5 \\
PmNiO$_2$ & -53.2 & -42.4 & -58.3 & -86.8 \\
SmNiO$_2$ & -54.2 & -43.2 & -55.4 & -82.5 \\
EuNiO$_2$ & -55.5 & -42.3 & -52.8 & -78.3 \\
GdNiO$_2$ & -56.1 & -41.5 & -50.7 & -75.6 \\
TbNiO$_2$ & -57.6 & -40.9 & -48.5 & -72.2 \\
DyNiO$_2$ & -58.6 & -40.0 & -46.6 & -69.3 \\
HoNiO$_2$ & -59.6 & -39.5 & -44.9 & -66.6 \\
ErNiO$_2$ & -60.7 & -38.5 & -43.2 & -64.0 \\
TmNiO$_2$ & -61.6 & -37.9 & -41.9 & -61.9 \\
YbNiO$_2$ & -62.5 & -36.7 & -40.2 & -59.4 \\
LuNiO$_2$ & -63.5 & -36.2 & -39.2 & -57.8 \\
\end{tabular}
\end{ruledtabular}
\end{table}
In the high–symmetry ${P4/mmm}$ phase, the Ni--O--Ni bond angle is constrained
to $180^{\circ}$. Decreasing the rare–earth ionic radius reduces the in–plane
lattice constant, thereby shortening both the Ni--Ni and Ni--O bond
lengths. The enhanced $d_{x^2-y^2}$–$p_{\sigma}$ hybridization increases the
effective hopping amplitude $t$, strengthening the antiferromagnetic
super-exchange interaction ($J_{\text{in}} \propto
\nicefrac{-t^2}{U}$). Consequently, $J_{\text{in}}$ increases systematically
across the series as the Ni--O bond length decreases. As discussed in the
previous section, however, this bond compression renders the ${P4/mmm}$
structure dynamically unstable for the smaller rare–earth members.
In the  ${I4/mcm}$ phase, the lattice strain is partially relieved
through an $a^{0}a^{0}c^{-}$ rotation of the NiO$_4$ squares. This distortion
maintains an approximately constant Ni--O bond length (Fig.~\ref{fig:b}) while
reducing the Ni--O--Ni bond angle below $180^{\circ}$. According to the
Goodenough–Kanamori–Anderson framework \cite{goodenough1955theory,
  kanamori1959superexchange, anderson1950antiferromagnetism}, the AFM
super-exchange interaction is maximized for a linear bond and decreases as the
bond angle deviates from $180^{\circ}$, due to the reduction of orbital
overlap and effective hopping. Consistently, the calculated $J_{\text{in}}$
values in the ${I4/mcm}$ phase are systematically smaller than those obtained
for the corresponding ${P4/mmm}$ structures of the same rare–earth ion.

Within the ${I4/mcm}$ series, the value of $J_{\text{in}}$ would depend on
both the rare–earth size, Ni--O--Ni bond angle and length. To understand the
intrinsic dependence of the in-plane magnetic coupling $J_{\text{in}}$ on the
rare-earth ionic size, we analyzed the quantity
$\log\left(\nicefrac{J_{\text{in}}}{\cos^2(180^\circ - \theta)}\right)$, where
$\theta$ is the Ni--O--Ni bond angle. This accounts for the angular dependence
of the super-exchange interaction, which scales with
$\cos^2(180^\circ - \theta)$ according to the Goodenough-Kanamori–Anderson
rules. Plotting this quantity against the rare-earth ionic radii yields a
three distinct linear trend (inline with observation made in
Fig.~\ref{fig:a}-\ref{fig:c}), indicating that the primary driver of
$J_{\text{in}}$ is the rare-earth ionic size, after
accounting for the geometric effects of the Ni--O--Ni bond angle and length.

\begin{figure}[htbp]
\includegraphics[scale=0.9]{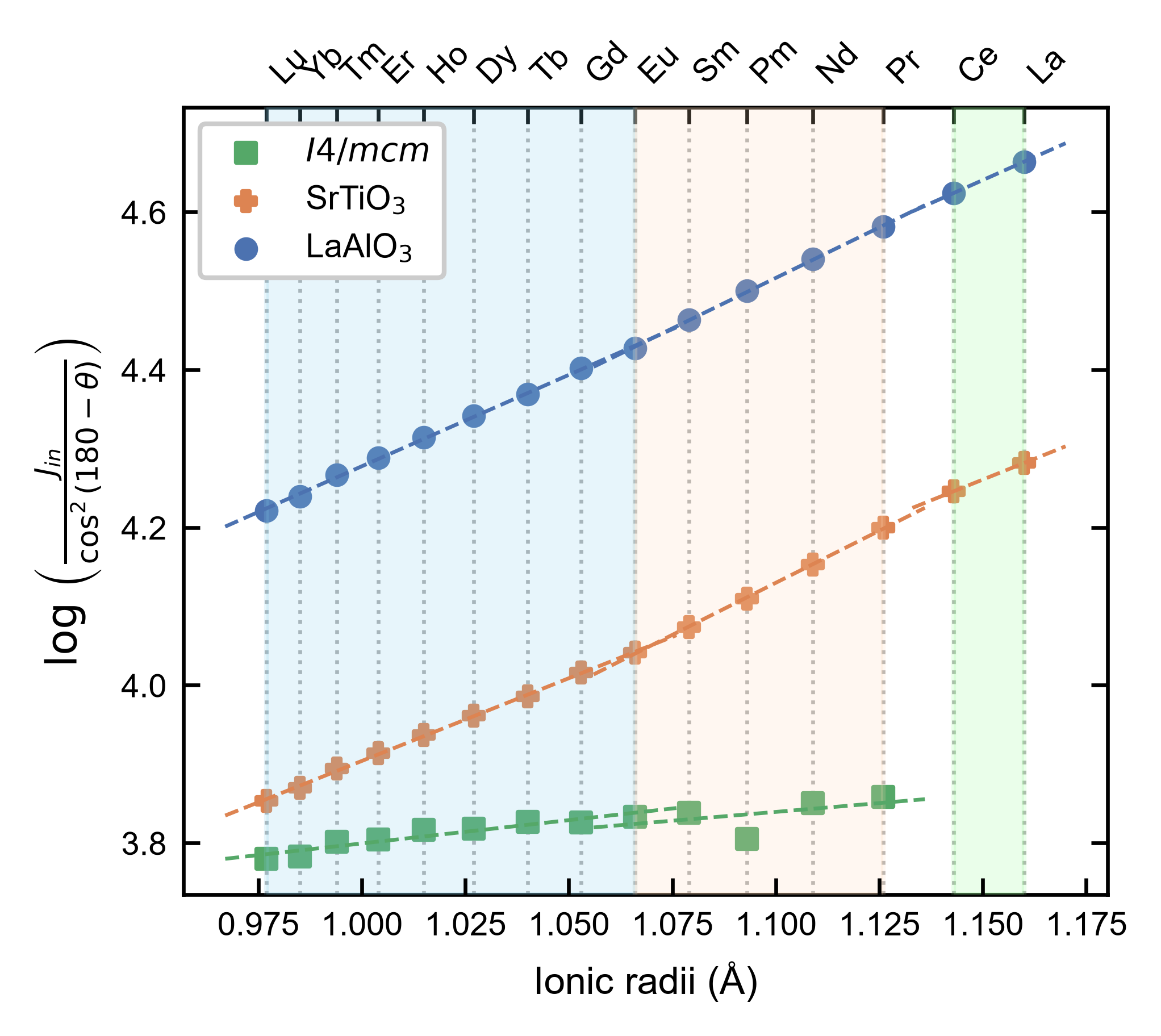}
\caption{Log of magnetic exchange interactions ($J_{\text{in}}$)
  over the cosine square of the Ni--O--Ni angle,
  plotted as a function of the rare
    earth ionic radius, in the bulk $I4/mcm$ phase  and in substrate-strained RNiO$_2$ systems
  ($R=$ La-Lu). Linear fits (dashed lines) are applied to three subspaces
  indicated by different shade colors.}
\label{fig:JVal-I4}
\end{figure}
Under epitaxial strain, the in-plane lattice constant is constrained by the
substrate. For a given rare-earth ion, the rotation amplitude is reduced
relative to the fully relaxed bulk ${I4/mcm}$ structure, bringing the
Ni--O--Ni bond angle closer to $180^\circ$. This enhances the effective
hopping and increases the magnitude of $J_{\text{in}}$ compared to the
untrained ${I4/mcm}$ case (Fig.~\ref{fig:JVal-I4}). The computed exchange
constants under SrTiO$_3$ strain for NdNiO$_2$ and PrNiO$_2$ (\unit[-61.8]{meV}
and \unit[-65.9]{meV}, respectively) are in good agreement with RIXS
measurements of
references~\onlinecite{lu2021magnetic,gao2024magnetic,PhysRevB.109.024512},
which report values of \unit[-63.6]{meV} and \unit[-66.5,-66.9]{meV}.

At this stage, we would like to draw the reader's attention to the closely
related quantum-chemical study of Katukuri \textit{et al.}~\cite{katukuri},
that obtained \unit[$J=-77.6$]{meV} for NdNiO$_2$ in the experimental $P4/mmm$
geometry of Ref.~\onlinecite{HAYWARD2003839}, using a large-CASSCF~+~MRLCC
method on an embedded fragment. This value is appreciably larger than the
$J=\unit[-59.05]{meV}$ obtained with our method on the same geometry. Several
methodological differences may contribute to this discrepancy. Katukuri
\textit{et al.}  employed closed-shell Cu$^+$ total-ion pseudopotentials for
the neighboring Ni$^{1+}$ ions, whereas we employ self-consistently optimized
Ni$^{1+}$ TIPs.  The method used by Katukuri \textit{et al.} is focused on
best computing the total energy of each state (large state-specific CASSCF
complemented by MRLCC), then extract the $J$ value from the energy
difference. In the DDCI method, we rather focus on computing the most accurate
energy difference between the singlet and triplet state and thus the most
accurate $J$ value, (i) by avoiding state-specific orbital CASSCF biases that
can affect the energy differences in an uncontrolled way, (ii) by excluding
effect that bring the same energy shift to the singlet and triplet. This
methodology allows to better control the quality of the computed energy
difference, and thus $J$ value, with similar computational effort.

For LaNiO$_2$ grown on SrTiO$_3$, the calculated value (\unit[-72.4]{meV})
overestimates the experimental result
(\unit[-61.6]{meV})~\cite{PhysRevB.109.024512}. It should, however, be noted
that the LaNiO$_2$ compound used in Ref.~\cite{PhysRevB.109.024512} exhibit
evidence of charge ordering (unlike the Nd and Pr systems), whereas
charge-ordering was not possible in our DFT calculations due to the unit cell
size and imposed space group. Since charge ordering can significantly affect
magnetic exchange interactions, its absence in the computed structure may be
responsible for the discrepancies between the calculated and experimental
exchange couplings. Let us note that the origin and intrinsic nature of the
reported charge ordering remain a subject of debate.

For a given substrate,
$J_{\text{in}}$ decreases with decreasing rare–earth ionic radius, consistent
with the systematic increase in bond–angle deviation from \unit[180]{\!\degree} across the
series. Overall, these results demonstrate that the magnetic exchange
interaction is primarily governed by rare–earth ionic radii and Ni--O--Ni bond
angle.

\begin{figure}[htbp]
\includegraphics[scale=0.9]{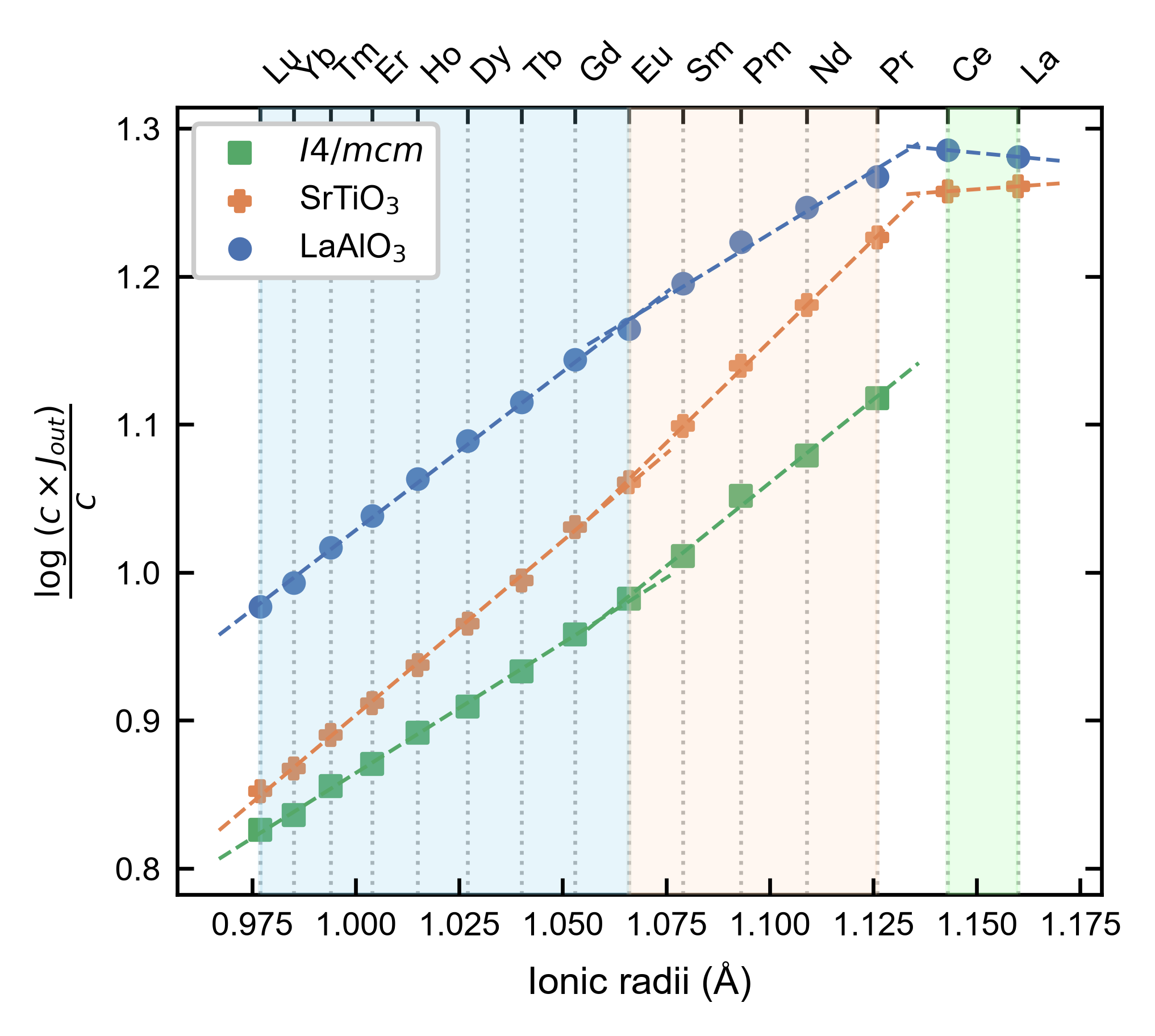}
\caption{Log of the magnetic exchange interaction by the
    out-of-plane lattice parameter ($J_{\text{out}}\times c$) ,
  normalized by the $c$ lattice parameter is plotted as a function of
  rare-earth ionic radii. Linear fits (dashed lines) are applied to three
  subspaces indicated by different shade.}
\label{fig:Jout}
\end{figure}

\begin{table}[htbp]
\caption{Out-of-plane magnetic exchange interaction $J_{\text{out}}$ (in meV)
    obtained from \textit{ab-initio} calculations for the RNiO$_2$ family.
    Different space groups and epitaxial strains on SrTiO$_3$ and LaAlO$_3$ substrates are reported.}
\label{tab:tableJout}
\begin{ruledtabular}
  \begin{tabular}{ccccc}
    System & \multicolumn{2}{c}{Bulk phase} & \multicolumn{2}{c}{Thin films} \\
  & ${P4/mmm}$ & ${I/4mcm}$ & SrTiO$_3$ & LaAlO$_3$ \\
 &  &  &  &  \\
\hline
LaNiO$_2$ & -17.22 & - & -22.29 & -25.06 \\
CeNiO$_2$ & -16.86 & - & -21.15 & -24.32 \\
PrNiO$_2$ & -16.53 & -12.69 & -18.53 & -22.29 \\
NdNiO$_2$ & -16.24 & -10.99 & -15.65 & -20.27 \\
PmNiO$_2$ & -15.92 & -9.88 & -13.34 & -18.25 \\
SmNiO$_2$ & -15.67 & -8.55 & -11.51 & -16.33 \\
EuNiO$_2$ & -15.53 & -7.69 & -10.04 & -14.54 \\
GdNiO$_2$ & -15.23 & -7.04 & -8.99 & -13.37 \\
TbNiO$_2$ & -14.99 & -6.41 & -7.87 & -11.92 \\
DyNiO$_2$ & -14.79 & -5.89 & -7.08 & -10.79 \\
HoNiO$_2$ & -14.60 & -5.50 & -6.40 & -9.79 \\
ErNiO$_2$ & -14.42 & -5.10 & -5.83 & -8.93 \\
TmNiO$_2$ & -14.26 & -4.82 & -5.40 & -8.24 \\
YbNiO$_2$ & -14.09 & -4.50 & -4.98 & -7.55 \\
LuNiO$_2$ & -14.01 & -4.33 & -4.72 & -7.13 \\
\end{tabular}
\end{ruledtabular}
\end{table}
Recently some authors found a non negligible out-of-plane magnetic
interaction~\cite{rosa2025spin}. This is a surprising result as it is
generally accepted in the literature that there is no long-range magnetic
order. We thus decided to compute the magnetic exchange interactions along the
out-of-plane direction, $J_{\text{out}}$. The $J_{\text{out}}$ values as a
function of the substrate and R are summarized in
Table~\ref{tab:tableJout}. Along the out-of-plane direction, there are no
ligand atoms, so the exchange coupling arises purely from Ni--Ni interactions
given by the FM direct exchange ($J_{d}$) and the kinetic AFM exchange
($J_{ke}$) terms. Given that the magnetic orbitals are predominantly of
$d_{x^2-y^2}$ character, the interaction should be of $\delta$--type, leading
to a reduced overlap term. This is consistent with our calculations, where we
found that the direct exchange term ($J_{d}$) is significantly reduced and
close to zero. Therefore, $J_{\text{out}}$ is proportional to the
antiferromagnetic kinetic exchange term ($\nicefrac{-t^2}{U}$). The magnetic
exchange coupling $J_{\text{out}}$ should decay exponentially with the Ni--Ni
distance along the out-of-plane direction,
$J_{out} \propto \nicefrac{e^{-\alpha R}}{R}$ where R corresponds to the
lattice parameter $c$. To understand how the rare-earth ionic radii affect the
$J_{\text{out}}$ term, we plot
$\nicefrac{\log \left(c\times J_{out}\right)}{c}$ against the ionic radii in
Fig.~\ref{fig:Jout}. For small ionic radii, the trend is linear, indicating
that $J_{\text{out}}$ decays exponentially with the Ni--Ni distance, as
expected for kinetic exchange. However, for larger ionic radii (Ce- and
La-based system), as they adopt a ${P4/mmm}$ geometry, additional factors such
as saturation of the Ni--Ni distance or rare-earth-specific effects may come
into play, leading to a different behavior. In conclusion, like Rossa {\it et
  al}.~\cite{rosa2025spin} we found substantial out-of-plane magnetic
interactions. However our results cannot be quantitatively compared to theirs,
as the discrepancies in the $J_{\text{in}}$ values between their RIXS
measurements on one side, and both the RIXS measurements
Ref.~\onlinecite{gao2024magnetic,PhysRevB.109.024512} and our calculations on
the other side, is large enough to suspect that their sample present some
significant compositional/structural differences with the later.  However, the
predicted out-of-plane AFM ordering is consistent with the experimental
observations of Ref.~\cite{rosa2025spin}.

The comparison between strained and unstained systems further highlights the
important role of epitaxial strain. Superconductivity in infinite–layer
nickelates has so far been realized only in thin films, whereas bulk
counterparts remain non-superconducting. The parent compounds of
unconventional high–temperature superconductors, such as cuprates and iron
pnictides, are characterized by strong AFM exchange
interactions, which are widely believed to provide the pairing
interaction. Although the exchange constants in nickelates remain smaller than
those of cuprates (typically \unit[-120 to -180]{meV})~\cite{Peng2017}, the substantial enhancement of
both $J_{\text{in}}$ and $J_{\text{out}}$ under strain indicates that
epitaxial constraints strengthen antiferromagnetic correlations. The increase
in $J$ under compressive epitaxial strain is qualitatively consistent with
experimental observations of enhanced superconducting properties in strained
nickelate films. In particular, a $\sim 40\%$ enhancement of the
superconducting critical temperature has been reported for
Pr$_{0.8}$Sr$_{0.2}$NiO$_2$ films grown on LSAT substrates (\unit[$a = 5.470$]{\AA})
compared to those grown on SrTiO$_3$. In our calculations, a systematic
increase in $J_{\text{in}}$ and $J_{\text{out}}$ are also observed with
increasing compressive strain, reflecting the strengthening of the
super-exchange interaction as the Ni--O--Ni bond angle approaches \unit[180]{\!\degree}.

It should however be noted that excessively large epitaxial strain may
introduce structural defects in nickelate thin films. For example, recent
studies on the Ruddlesden–Popper compound Nd$_4$Ni$_3$O$_{10}$ have shown that
epitaxial strain can generate extended structural defects depending on the
strain state~\cite{ferenc2023limits}. Films reduced under compressive strain
on substrates such as LaAlO$_3$ were found to develop $c$-axis canting defects
and structural disorder, whereas growth under large tensile strain can lead to
a high density of rock–salt fault defects. Consequently, although compressive
strain can enhance the magnetic exchange interaction and potentially favor
superconductivity, this effect is expected to hold only within a regime where
the film remains structurally coherent. These observations highlight the
central role of substrate–induced structural tuning, as well as the
limitations imposed by strain–induced defects, in governing magnetic
interactions and their possible connection to superconductivity in
infinite–layer nickelates.

\section{Conclusion}

In this work, the nearest–neighbor magnetic exchange interaction in RNiO$_2$
(R $=$ La--Lu) was investigated as a function of rare–earth size and epitaxial
strain using {\it ab-initio} multireference configuration interaction
calculations. The crystal structures were first optimized using hybrid
functional DFT simulations. For the untrained compounds, the ground state
structure was found to adopt the ${I4/mcm}$ symmetry (except for R = La and
Ce), consistent with recent theoretical predictions~\cite{philipghosez,xia,carrasco}. This
phase is characterized by a rotation of the NiO$_4$ plaquettes around the
out-of-plane axis. While the lattice parameters decrease systematically with
decreasing rare-earth ionic radius, the amplitude of this rotational
distortion increases along the series.

The effect of epitaxial strain was further examined by imposing substrate
lattice parameters. Under compressive strain, the rotation angle of the
NiO$_4$ plaquettes is reduced, while the rare-earth–oxygen (R--O) distance
remains nearly unchanged across different strain conditions, effectively
acting as a structural constraint in the system.

The computed magnetic exchange interactions were found to be strongly
dependent on the optimized crystal structure. Compared to the high-symmetry
${P4/mmm}$ phase, the exchange interaction is reduced in the 
${I4/mcm}$ structure due to the deviation of the Ni--O--Ni bond angle from
linearity. Upon the application of compressive strain, the exchange
interaction increases for a given rare-earth compound, as the Ni--Ni distance
decreases and the Ni--O--Ni bond angle approaches towards $180^\circ$, thereby
strengthening the superexchange pathway in $J_{\text{in}}$.

Overall, the calculated values of $J_{\text{in}}$ decrease with decreasing
rare-earth ionic radius and increase under compressive strain. The values
obtained for RNiO$_2$ under SrTiO$_3$ substrate strain are in good agreement
with experimentally estimated exchange energies from RIXS measurements of
Ref.~\onlinecite{lu2021magnetic, gao2024magnetic,PhysRevB.109.024512}. These
results highlight the important role of substrate-induced structural tuning in
modifying magnetic interactions in infinite-layer nickelates and provide
microscopic insight into how epitaxial strain may influence the correlated
electronic properties of these materials. The out-of-plane magnetic exchange
coupling $J_{\text{out}}$ is dominated by kinetic exchange due to the
$d_{x^2-y^2}$ orbital character, leading to significantly reduced
ferromagnetic exchange coupling term. The finite out-of-plane exchange
interaction obtained in the present work, together with the recent RIXS
observation of a non-negligible interlayer exchange~\cite{rosa2025spin},
indicates that the magnetic interactions in RNiO$_2$ are inherently
three-dimensional. This makes the absence of experimentally established
long-range magnetic order an  intriguing open-question.

\section{Acknowledgments}

This work was financially supported by the ANR ACCURATE (Agence Nationale de la Recherche projet
N$^\circ$ANR-23-CE08-0017-02). The simulations were performed using IDRIS, CINES and
TGCC computing resources under GENCI project N$^\circ$0801842.

\bibliography{NiO2}

\end{document}